\documentclass[review, 12pt]{elsarticle}

\usepackage[utf8x]{inputenc}
\usepackage[english]{babel}
\usepackage{textcomp}
\usepackage{esdiff}
\usepackage[T2A]{fontenc}
\usepackage{amsmath}
\usepackage{graphicx}
\usepackage{wrapfig}
\makeatletter
\def\ps@pprintTitle{%
 \let\@oddhead\@empty
 \let\@evenhead\@empty
 \def\@oddfoot{}%
 \let\@evenfoot\@oddfoot}
\makeatother

\begin{document}
\begin{frontmatter}
\title{Simulation of magnetodielectric effect in magnetorheological elastomers}
\author[1]{D.A. Isaev}
\ead{isaev.danil@gmail.com}
\author[1,2]{A.S. Semisalova}
\ead{semisalova@physics.msu.ru}
\author[1]{Yu.A. Alekhina}
\ead{ya.alekhina@physics.msu.ru}
\author[1,3]{L.A. Loginova}
\ead{la.loginova@physics.msu.ru}
\author[1,3]{N.S. Perov}
\ead{perov@magn.ru}
\address[1]{Lomonosov Moscow State University, Faculty of Physics, Moscow, Russia}
\address[2]{Helmholtz-Zentrum Dresden - Rossendorf, Institute of Ion Beam Physics and Materials Research, Dresden, Germany}
\address[3]{Baltic Federal University, Kaliningrad, Russia}

\begin{abstract}
We present the results of numerical simulation of magnetodielectric effect (MDE) in magnetorheological elastomers (MRE) – the change of effective permittivity of elastomer placed under the external magnetic field. The computer model of effect is based on the assumption about the displacement of magnetic particles inside the elastic matrix under the external magnetic field and the formation of chain-like structures. Such displacement of metallic particles between the planes of capacitor leads to the change of capacity, which can be considered as a change of effective permittivity of elastomer caused by magnetic field (magnetodielectric effect).
In the literature mainly 2D approach was used to model similar effect. In this paper we present the new approach of magnetorheological elastomers simulation -  3D-model of magnetodielectric effect with ability to simulate systems of ~$10^6$ particles.
Within the framework of the model three types of particle size distributions were simulated which gives an advantage over previously reported approaches. Lognormal size distribution was shown to give better qualitative match of the modeling and experimental results than monosized type.
The developed model resulted in perfect qualitative agreement with all experimental data obtained earlier for Fe-based elastomers. The proposed model is useful to study these novel functional materials, analyze the features of magnetodielectric effect and predict the optimal composition of magnetorheological elastomers for further profound experimental study.
\end{abstract}
\end{frontmatter}

\section{Introduction}
Magnetorheological materials due to variety of their properties belong to the class of so called smart materials. They can respond to the changes in their environment and actively alter some physical properties according to these changes. Such responsiveness defines practical importance of these materials and their wide use in different fields from everyday exploitation to industrial applications \cite{007, 008, 044, 045, 046, 047}.\par
Magnetorheological elastomers (MRE) usually consist of soft polymer matrix with magnetic particles dispersed in it. Recent experimental studies show that MREs exhibit the mechanical and electrical properties changes under the external magnetic field: magnetorheological effect \cite{018}, magnetoresistance effect \cite{043}, magnetodielectric effect \cite{Semisalova13}, etc. Change of properties of MRE may be explained by change of the particle distribution inside of the matrix. Particles tend to concentrate to column-like structures along the direction of the external magnetic field causing the shape deformation of the sample. This also leads to the anisotropy of the effects.\par
Bica et al. have studied the variation of capacitance of capacitor filled with MRE which was dependent on the distance between the plates \cite{017, 035}, the concentration dependence of this effect \cite{033}, its anisotropy \cite{034},  electroconductive properties \cite{032}. Miao Yu et al studied wave-absorbing properties of MRE \cite{040} and conductivity of magnetorheological gels \cite{041}. Knicht et al introduced model of contact resistance and studied piezoresistivity of MRE \cite{037}. In \cite{038} they extend their model for thermoresistance and show the decrease of resistance of MRE with the external field increase. Wang Yu et. al. studied the fatigue dependences of properties of MRE \cite{039}. It was noted that capacitance tends to increase for the same value of magnetic field for more “old” materials because of breaking of its inner structure. Semisalova et al \cite{Semisalova13} have reported the significant change of effective permittivity of MRE measured using plane capacitor with fixed distance between the plates. Kramarenko et al studied hysteresis characteristics of MDE in magnetoactive materials \cite{036}.
There are the number of different approaches to the theoretical estimation of such effects. Homogenization and estimation through so called effective medium may be used for mechanical properties. Electrical properties may be estimated through an influence of the single regular chain of particles or the imagine filler as cuboids with changing linear dimensions.\par
In this work we present a model and results of the direct computer numerical modelling of the magnetodielectric effect in MRE – the change of capacity of a plane capacitor filled with magnetic elastomer and placed under the external magnetic field.
\section{Model}
We consider MRE as a system of spherical particles in the elastic matrix. This system is affected by the homogenous magnetic field. Each particle in such system is in the field of magnetic dipoles, the external magnetic field and in the elastic forces field. We assume that filler particles are the soft magnetic material and the external magnetic field can only orient magnetic moments of particles.  Therefore, the moments of particles are oriented towards effective field, which is the sum of the external field and nearest neighbours gradient field.
Particles move under influence of the gradient field generated by the nearest particles, and the field of elastic forces generated by the particle displacement in polymer matrix relative to their initial equilibrium position.
\subsection{Magnetic force}
We calculate gradient magnetic field on given particle as sum of dipole field of particle nearest neighbours:
\begin{equation}
\vec{H}_{nn}=\sum_{k=1}^{N}\vec{H}_k=\sum_{k=1}^{N}\frac{3(\vec{m}^k\vec{R}_k)-R_k^2\vec{m}^k}{R_k^5},
\end{equation}
where $\vec{m}^k$ is the magnetic moment of k-th neighbour, $R_k$ is the distance between the given and neighbour particle.
So for magnetic force $F^{magn}$ acting on the particle we obtain the following equation:
\begin{equation}
\vec{F}^{magn} = \nabla{\vec{H}_{nn}}\vec{m}
\end{equation}
Or for each projection: 
\begin{gather}
F^{magn}_{x_j}=\sum_{x_i}m_{x_i}\frac{dH_{x_i}}{dx_j}\\
\frac{dH_{x_i}}{dx_j}=\sum_{k=1}^{N}\frac{A_kR^2_k-5B_k(x_j-x_j^k)}{R^7_k},\\
A_k=3m_{x_j}^k(x_i-x_i^k)+3(\vec{m}^k\vec{R}_k)\delta_{ij}-2(x_j-x_j^k)m_{x_i}^k,\\
B_k=3(\vec{m}^k\vec{R}_k)(x_i-x_i^k)-R_k^2m_{x_i}^k
\end{gather}
where $x_i$ and $x_j$ are independent coordinate axes $(x, y, z)$, $\delta_{ij}$ is Kronecker delta, $\vec{m}$ magnetic moment of the particle, $x$ and $x_k$ are coordinates of particle and its k-th neighbour.
\subsection{Elastic force}
For elastic interaction of particles with the polymer matrix we use simple spring approach \cite{spring1,spring2}. Each particle is assumed to be in the field of elastic force which tends to return the particle to the initial position if the displacement happens.
By definition of elastic modulus $E$, where $F$ is normal part of acting force, $S$ is the area, l is length of deforming media and $\delta{l}$ is the displacement, we obtain:
\begin{equation}
E = \frac{F/S}{\delta{l}/l} = \frac{Fl}{S\delta{l}}
\end{equation}
We assume $S$ to be an area of a diametral cross-section of the spherical particle, and $l$ to be average distance between particles. For the elastic force $F^{elastic}$ acting on the particle we obtain:
\begin{equation}
F^{elastic} = 2\pi E\frac{r^2}{l}\delta{l}
\end{equation}
\subsection{Numerical method}
The equation of particle motion is given by:
\begin{equation}\label{eq:motion}
\begin {aligned}
m\frac{d^2\vec{r}(t)}{dt^2}=\vec{F}^{magn} + \vec{F}^{elastic}\\
\vec{r}(0)=\vec{r}_0\\
\frac{d\vec{r}(0)}{dt}=0.
\end{aligned}
\end{equation}
For each particle we set the initial position, the velocity of particles is set to zero. Particles are still at the initial state.
To solve this equation numerically we used the Verlet algorithm \cite{Verlet}. We approximate the function $r(t)$ in $h$-neighbourhood of $t$ as Taylor series:
\begin{equation}
\begin {aligned}
r(t+h)=r(t)+\diff{r}{t}h+\frac{1}{2}\frac{d^2r}{dt^2}h^2+\frac{1}{6}\frac{d^3r}{dt^3}h^3+O(h^4),\\
r(t-h)=r(t)-\diff{r}{t}h+\frac{1}{2}\frac{d^2r}{dt^2}h^2-\frac{1}{6}\frac{d^3r}{dt^3}h^3+O(h^4).
\end{aligned}
\end{equation}
After summation and substitution we get:
\begin{equation}
r(t+h)=2r(t)-r(t-h)+\frac{1}{m}F(t)h^2+O(h^4).
\end{equation}
Or assuming $h=t_{i+1}-t_i$:
\begin{equation}
r_{i+1}=2r_i-r_{i-1}+\frac{1}{m}F_ih^2+O(h^4).
\end{equation}
So the position of each particle at each iteration depends on two previous positions and does not depend on particle speed. This equation represents the Verlet integration. 
\subsection{Capacitance calculation}
\begin{figure}[!htb]
\centering
\includegraphics[width=0.5\textwidth]{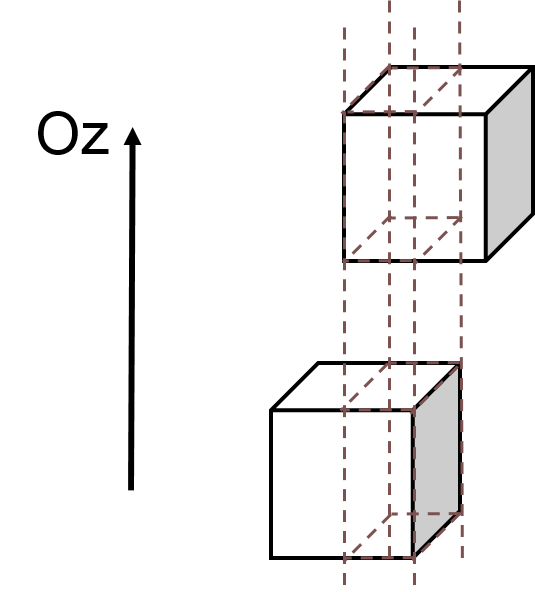}
\caption{Split the volume along Oz axis.}
\label{fig:oz}
\end{figure}
Important part of our model is the calculation of capacitance. We used area-cap approximation to solve this problem:
\begin{equation}
C = \epsilon_0\epsilon\frac{S}{d}
\end{equation}
It is pure geometrical approach. We assume, that electric field inside the capacitor is not changed and remains uniform and constant. In addition, we approximate spherical particles with axis-aligned shapes (cuboids). Therefore, we can imagine our sample as a system of interconnected planar capacitors. To determine an overall capacitance we split volume along the Z-axis as shown on figure \ref{fig:oz}, which is perpendicular to the capacitors plates, into areas, so that each particles either completely overlaps this area, or does not overlap at all. In this way, each slice along Z-axis is a series of capacitors, while the slices are connected in parallel. To obtain such partition we need to review projections of particles onto the plane. Therefore, problem is reduced to the analysis of two-dimensional geometry. To solve this problem the sweep-line algorithm was used. The sweep line is an imaginary line running along the X-axis in one direction. It processes “events” – in our case the event is the crossing with particle boundaries. After registration of two consecutive events, we get the narrow band at which second line starts running, but along the Y-axis. This way we get the needed geometry split and calculate overall capacitance. 
\begin{figure}[!htb]
\minipage{0.32\textwidth}
  \includegraphics[width=\linewidth]{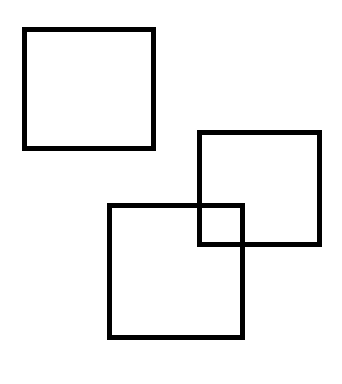}
\endminipage\hfill
\minipage{0.32\textwidth}
  \includegraphics[width=\linewidth]{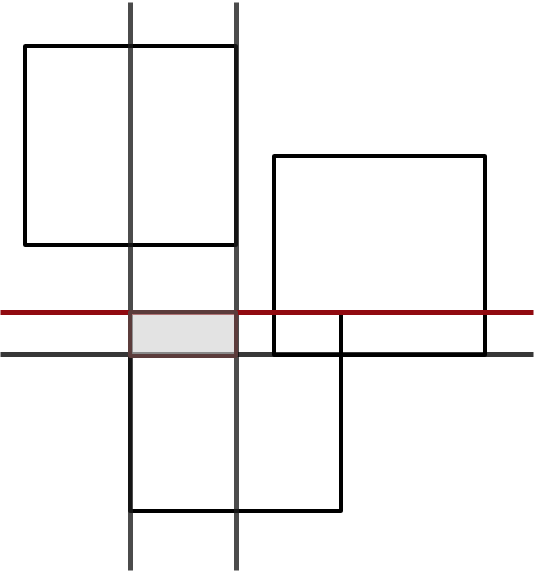}
\endminipage\hfill
\minipage{0.32\textwidth}
  \includegraphics[width=\linewidth]{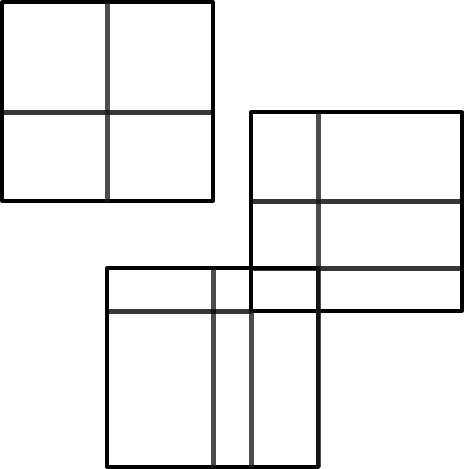}
\endminipage
\caption{Sweep line example. Left: Initial state (projections of particles). Middle: Sweep line cut area. Right: Final split}\label{fig:split_example}
\end{figure}

\subsection{Parameters}
We implemented algorithm using C++ programming language to solve (\ref{eq:motion}).
For all simulations standard personal computer was used. We used parameters similar to experimental in our simulations. We considered Fe as particles material. Particles had average radius $R_0$ of 5 µm. Average stiffness $E_0$ of matrix (Young's modulus) was 0.01 GPa. We did not use any periodical approximations and simulated sample as whole using 25000-130000 particles in each simulation. Volume concentration was varied upto 27\%. \par
One of advantages of this model is that it is simple in terms of calculating. This allowed us to solve (\ref{eq:motion}) for system of $10^6$ particles with capacitance evaluation (average modelling point) in less than a minute on the ordinary laptop.

\subsection{Particles size distributions}
\begin{wrapfigure}{L}{0.5\textwidth}
\centering
\includegraphics[width=0.4\textwidth]{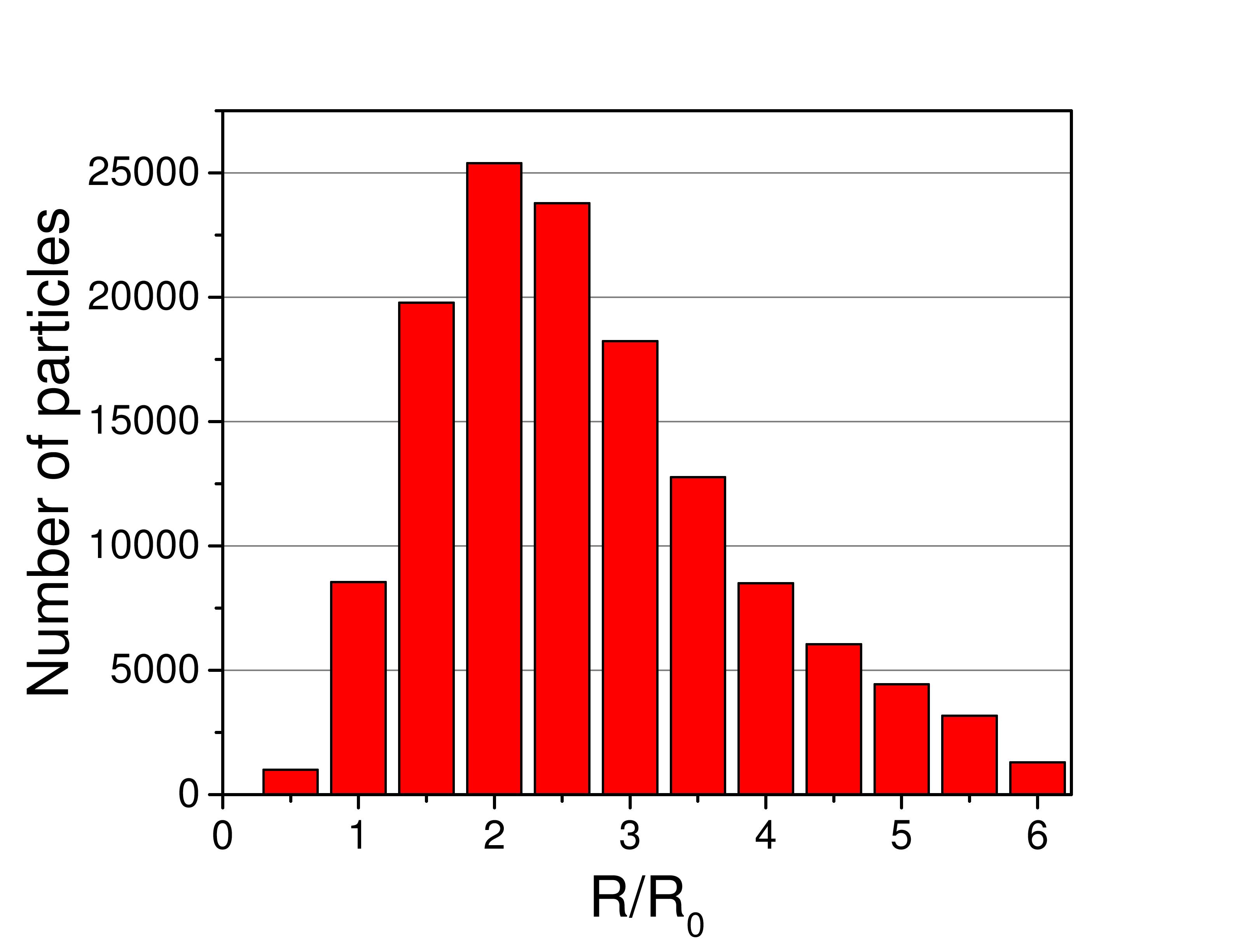}
\caption{Lognormal distribution of 130000 particles. $R_0$ corresponds to average radius of 5 µm.}
\label{fig:lognormal}
\end{wrapfigure}

We considered three types of particles size distributions. Monosized with all particles of the same size was used as a base type. Lognormal type was obtained by modulating monosized radius with lognormal distribution with median 1 and sigma 0.7. We also used bimodal as intermediate type with two types of particles size: small and big with three times volume differences between each other.

\section{Experiment}

Simulation data was compared with experimental data obtained by our group. Samples of magnetorheological elastomers were provided by G.V. Stepanov. Silicon Compound SIEL\textsuperscript{TM} produced by the Institute of Chemistry and Technology of Organoelement Compounds (GNIIChTEOS) was used as a matrix. Carbonil Iron particles with average diameter 5 µm were used as a ferromagnetic component. Synthesis of magnetorheological elastomers was described earlier \cite{048}\par
In experimental part samples with 40\%, 54\%, and 72\% mass fractions of carbonil iron were investigated. To compare with simulation results volume fractions were used. Corresponding volume fractions are 8.1\%, 13.9\% and 25.3\%.
Magnetodielectric effect was calculated from the values of capacitance of capacitor filled with elastomer in the absence and in presence of magnetic field  \cite{Semisalova13}.
Sample of magnetorheological elastomer with length 18 mm, width 10 mm and height 3mm was fixed between plates of capacitor and then placed between the poles of electromagnet so that the magnetic field was perpendicular to the capacitor plates. Capacitance was measured with RLC-meter AKTAKOM AM-3016 at zero magnetic field and under applied field with flux density 500mT. Field dependece of magnetization has been measured with vibrating sample magnetometer (VSM) Lake Shore 7407.\par

\section{Results and Discussion}
\subsection{Magnetic properties}
\begin{figure}[!htb]
\minipage{0.49\textwidth}
  \includegraphics[width=\linewidth]{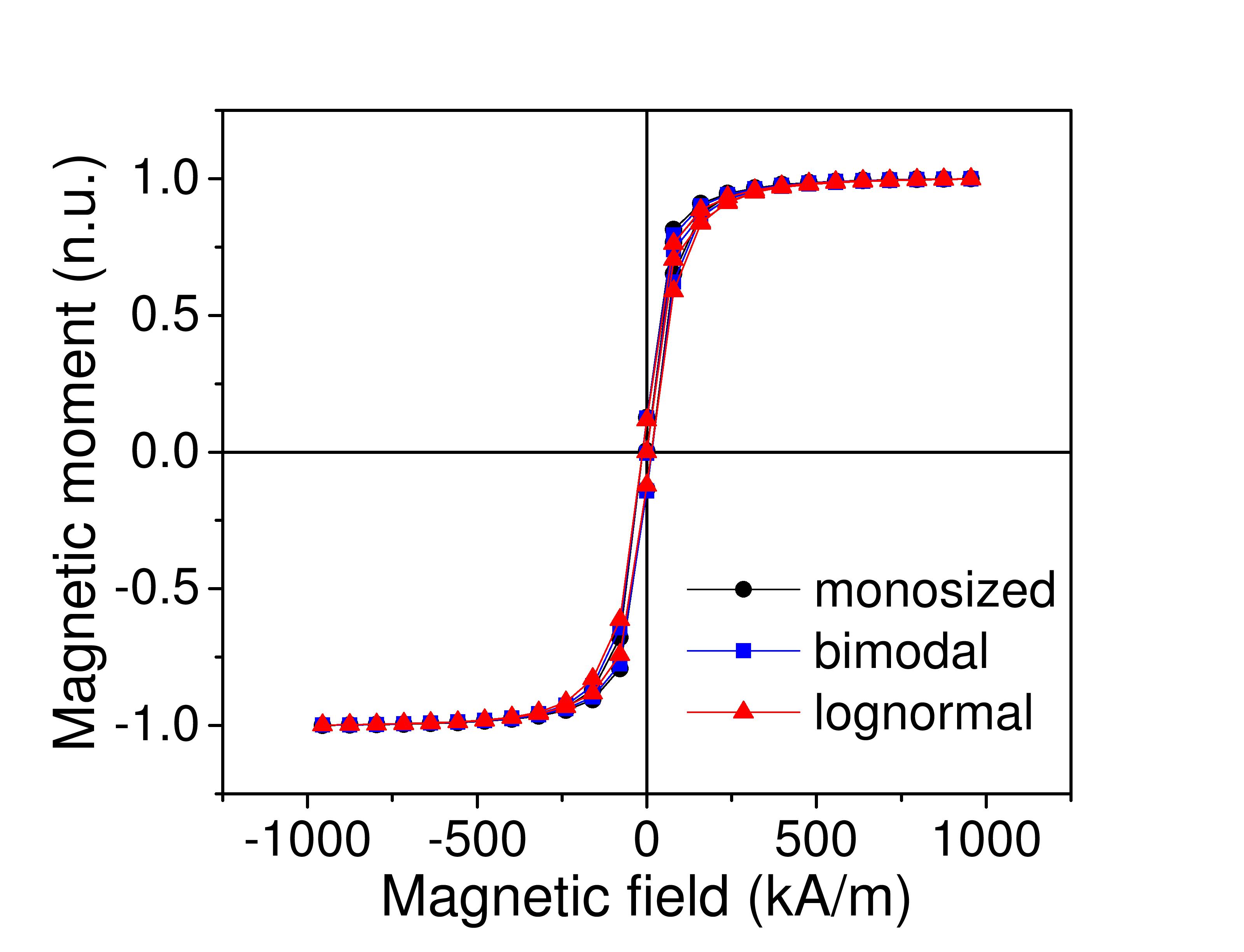}
  \caption{Simulated field dependence of magnetization of MRE with different types of size distribution of filler particles}\label{fig:hist_comp}
\endminipage\hfill
\minipage{0.49\textwidth}
  \includegraphics[width=\linewidth]{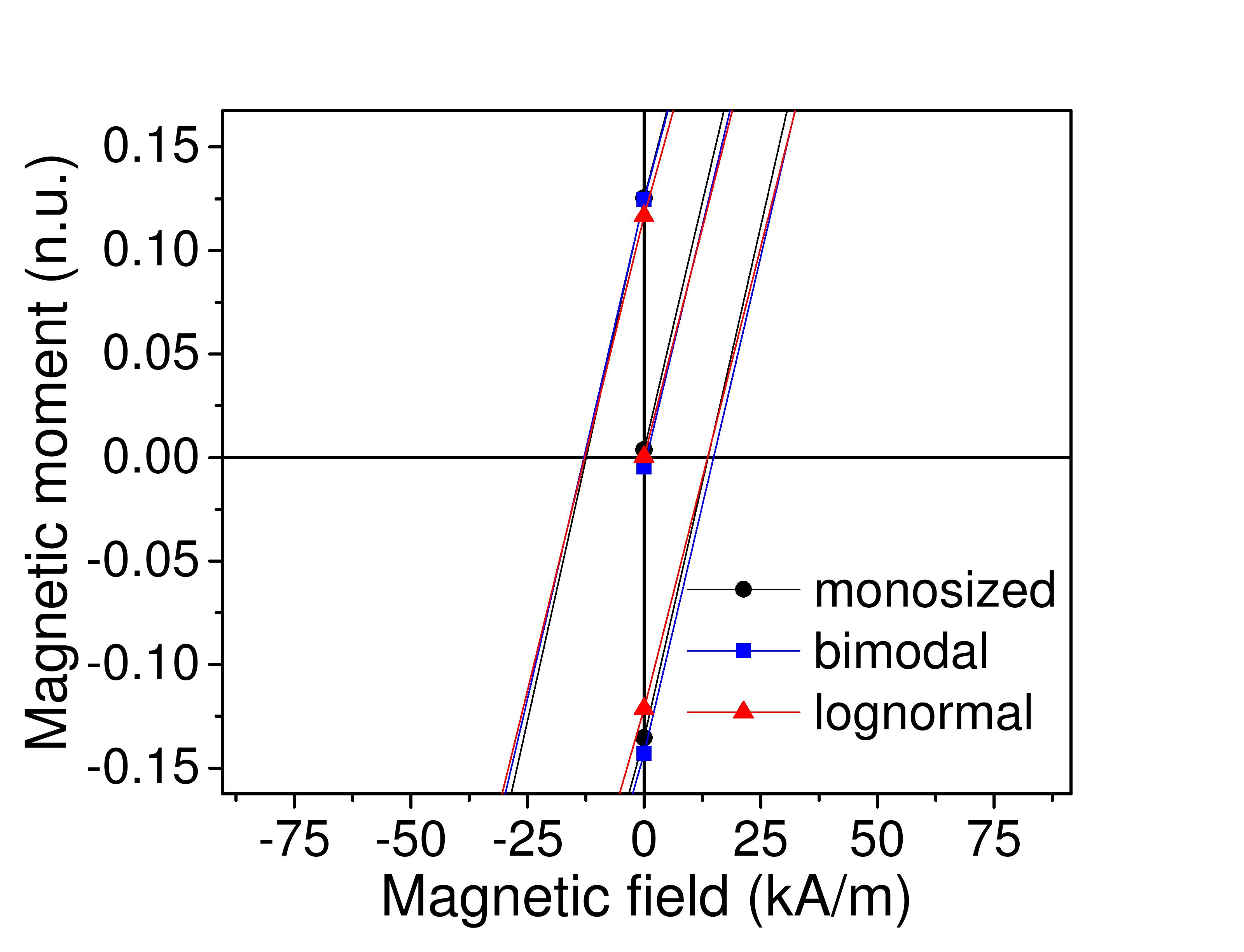}
  \caption{Low-field region showing the open hysteresis loop}\label{fig:coerc_force}
\endminipage\hfill
\end{figure}
\begin{figure}[!htb]
\includegraphics[width=\linewidth]{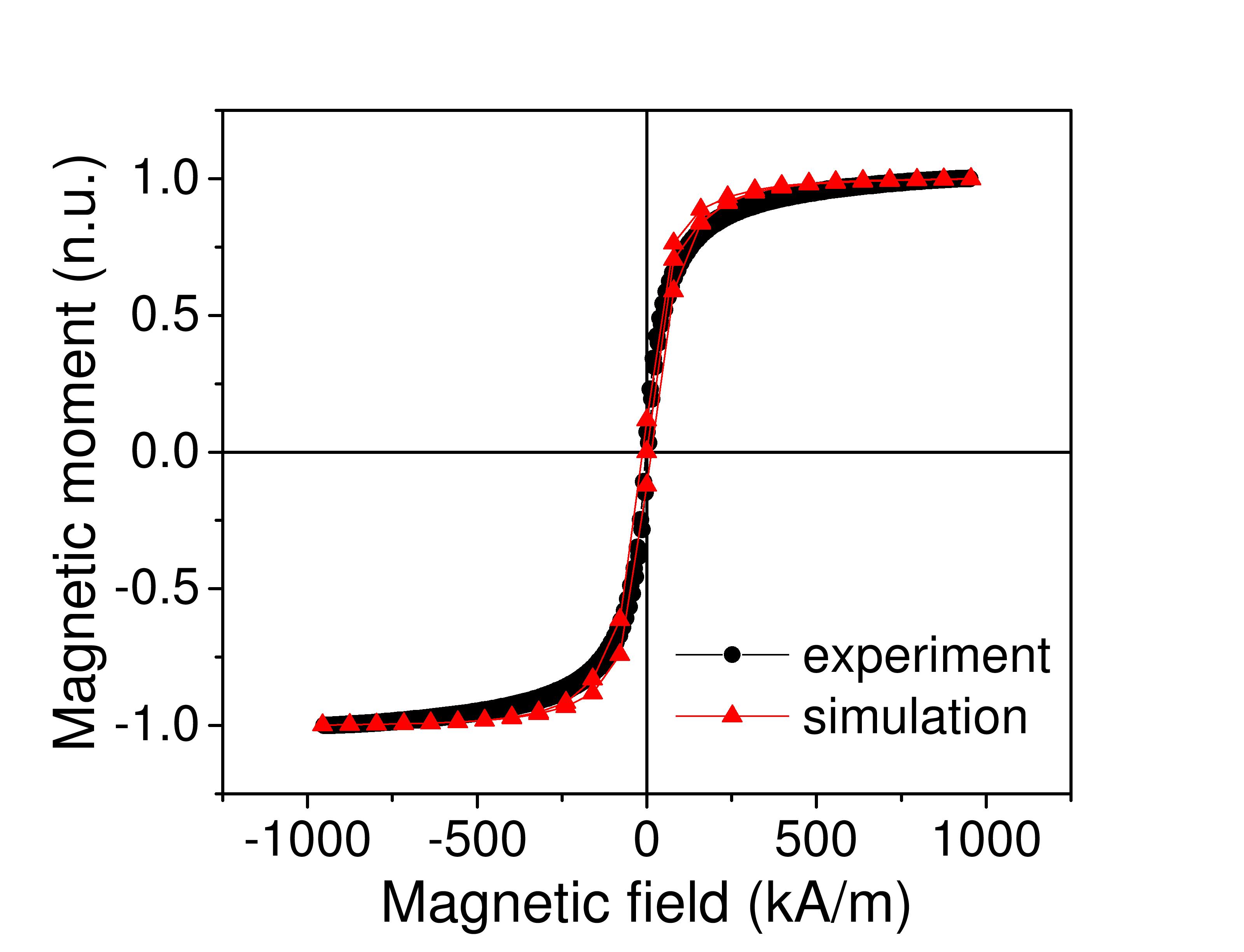}
\caption{The comparison of experimental and simulated magnetization of MRE vs. applied magnetic field. Experiment was performed for the MRE sample with 8.1\% volume concentration of Fe. Simulation was performed for the sample with 8.1\% volume of Fe, lognormal size distribution}
\label{fig:hist_comp_exp_dep}
\end{figure}

First, we simulated magnetic properties of the material, particularly the magnetic field dependence of magnetization. Simulation results show that we get the same magnetic hysteresis for different types of the particles size distribution - monosized, bimodal and lognormal (Fig. \ref{fig:hist_comp}). Also simulation shows presence of coercive field for all three types of MRE (Fig. \ref{fig:coerc_force}). The simulated field dependence matches well with the experimental results. Fig. \ref{fig:hist_comp_exp_dep} shows the comparison of magnetic hysteresis curves obtained using our model and measured with VSM.\par
\subsection{Magnetodielectric effect}
\begin{figure}[!htb]
\includegraphics[width=\linewidth]{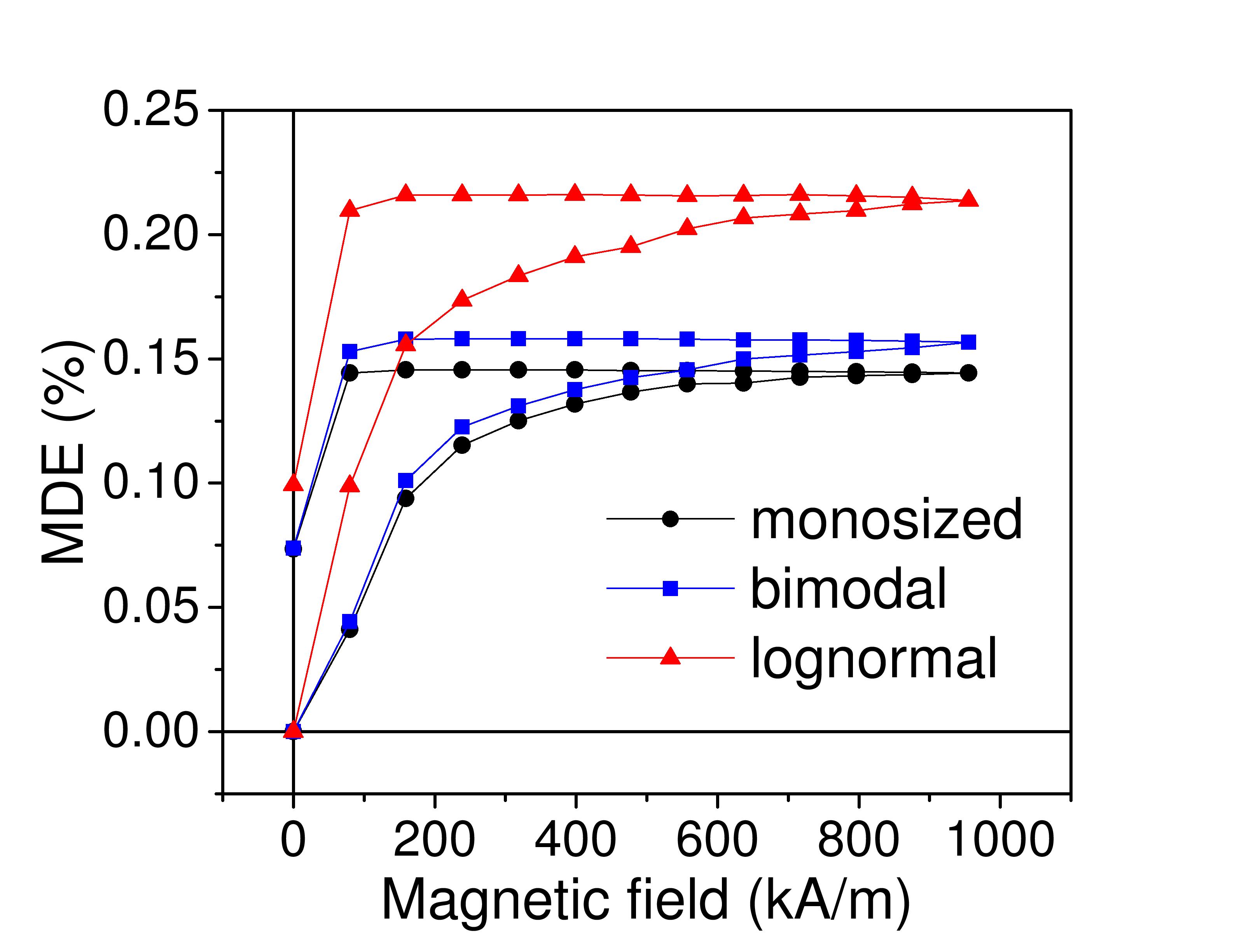}
\caption{Simulated field dependence of MDE for three different types of size distribution. Volume concentration of magnetic particles is 25\%; magnetic field is perpendicular to the capacitor plane.}
\label{fig:mde_comparison}
\end{figure}

The simulated field dependences of MDE in magnetic elastomers with three types of particles size distribution are shown on Fig. \ref{fig:mde_comparison}. All three types of samples show the same qualitative behaviour, hence they are different in the maximum MDE value, with lognormal distribution being the largest one. \par
\begin{figure}[!htb]
\minipage{0.49\textwidth}
  \includegraphics[width=\linewidth]{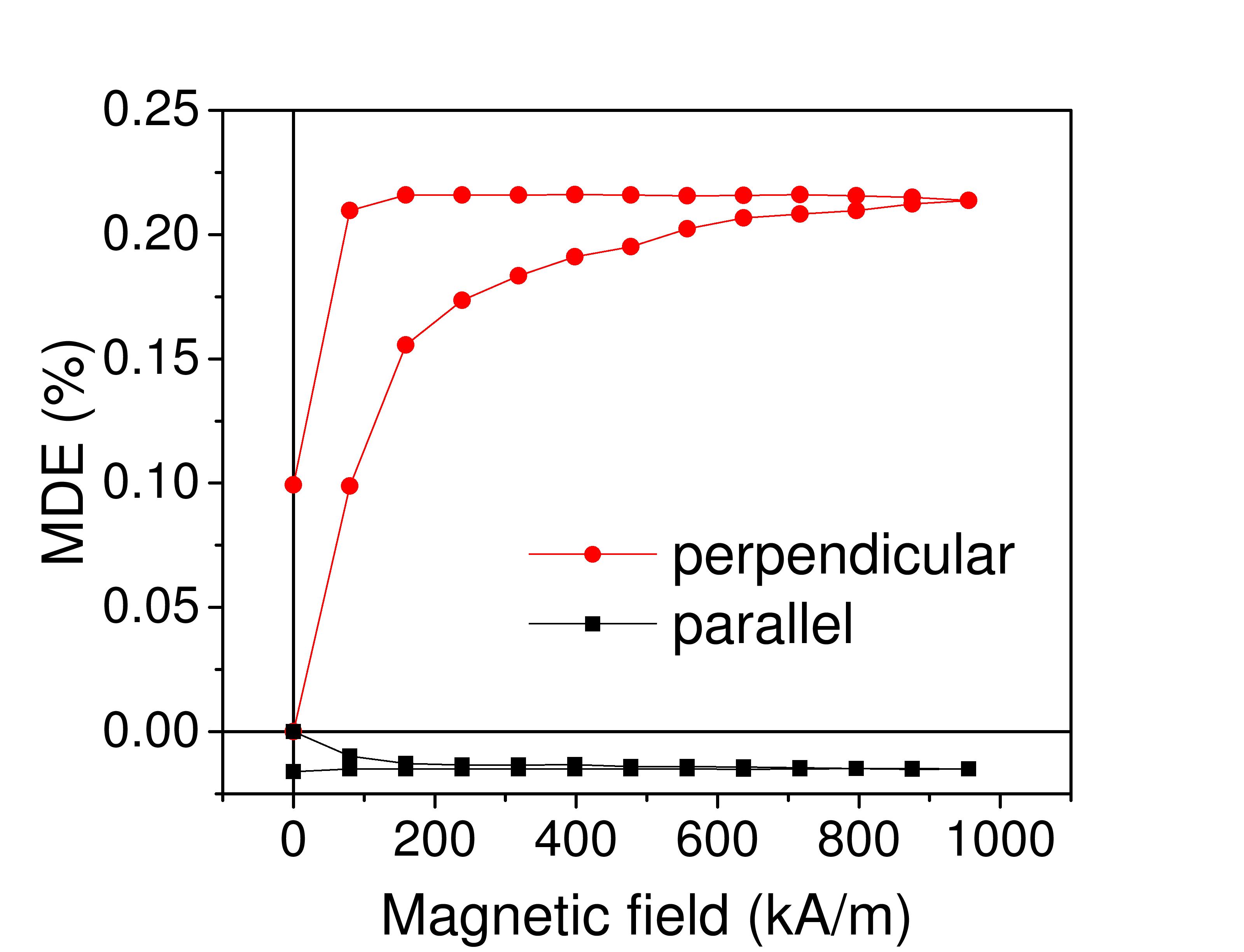}
\endminipage\hfill
\minipage{0.54\textwidth}
  \includegraphics[width=\linewidth]{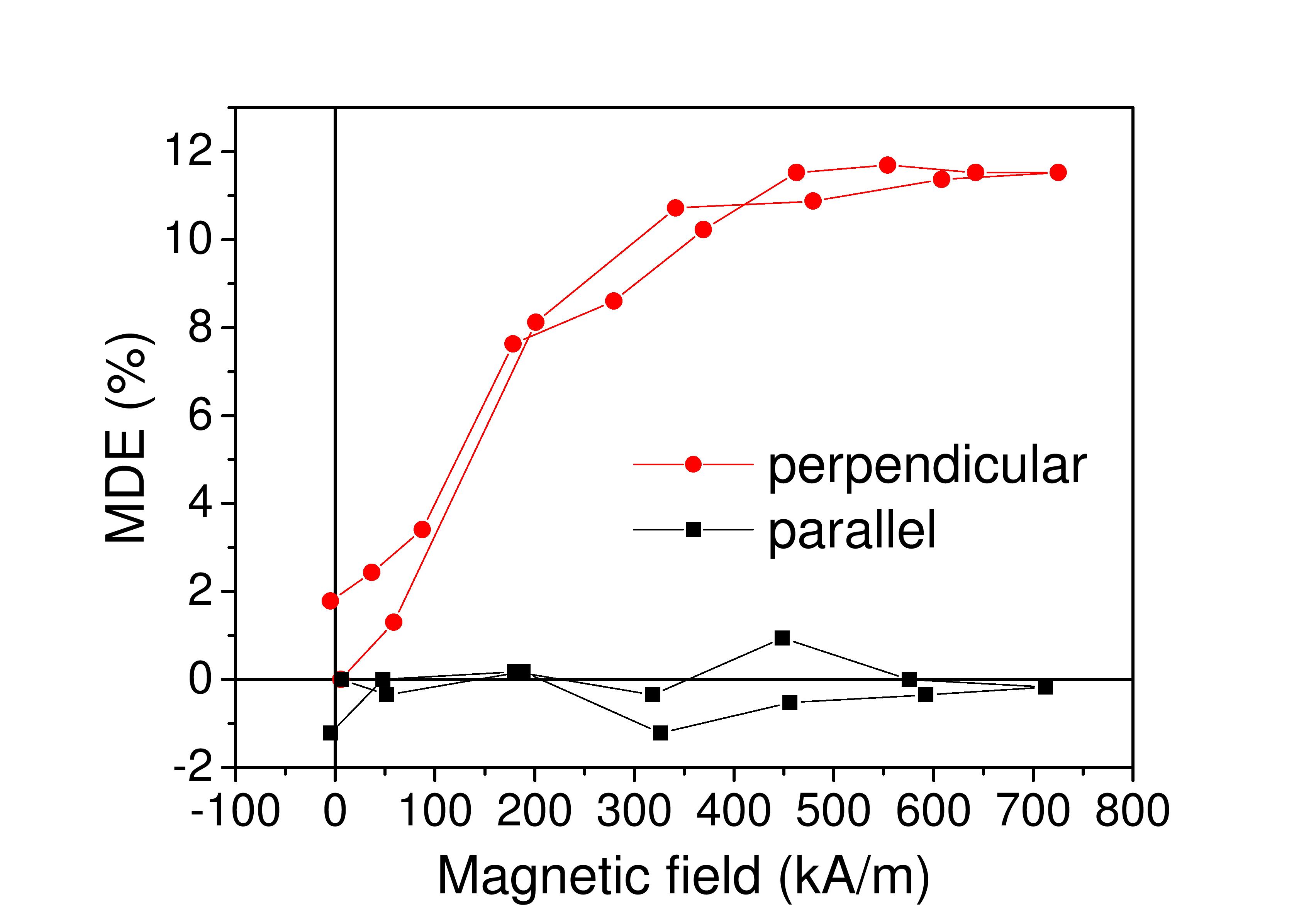}
\endminipage\hfill
  \caption{Comparison of simulated field dependence of MDE (left) with experimental data (right) for the magnetic field applied perpendicular and parallel to the capacitor plane. Simulation: sample with 25\% volume concentration of Fe, lognormal size distribution. Experiment: sample with 25.3\% volume concentration of Fe. Model shows good qualitative agreement.
}\label{fig:experiment}
\end{figure}

We compared simulation results for dependence of MDE on external magnetic field with experimental data reported in \cite{Semisalova13} and obtained for Fe-based MRE (Section 3). Comparison shows that the developed model describes the effect qualitatively well (Fig. \ref{fig:experiment}). The anisotropy of MDE was confirmed within our model, we have observed the difference in the effect sign and magnitude for different directions of the external field. Also, the hysteretic character of field dependence due to the polymer matrix elasticity was found similar to experimental observations. \par
\subsection{Concentration dependence of MDE}
\begin{figure}[!htb]
\includegraphics[width=\linewidth]{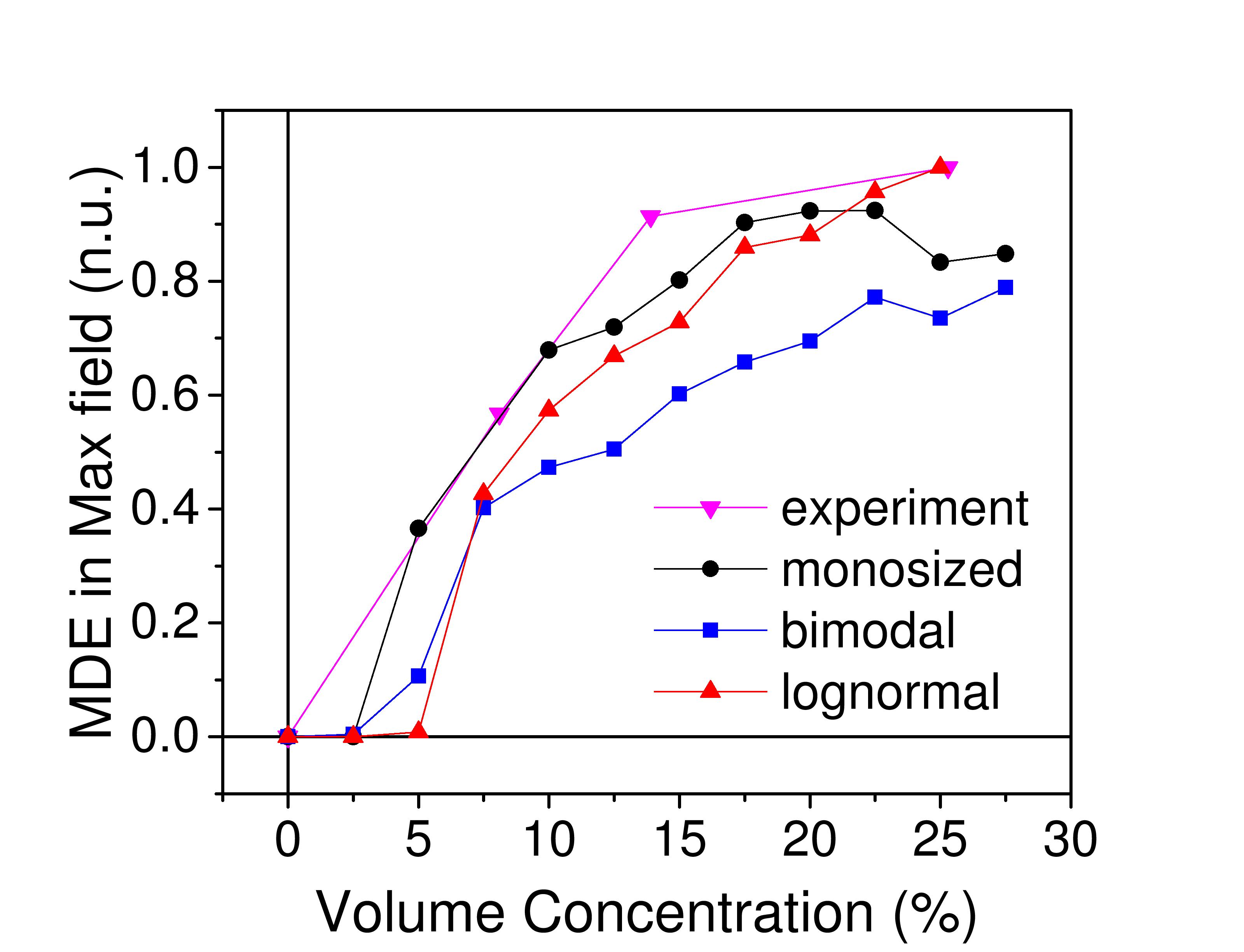}
\caption{Volume concentration dependence of maximal MDE value for three types of size distribution compared to experimental data.}
\label{fig:vol_conc_dep}
\end{figure}

MRE shows strong dependence of MDE on the magnetic particles mass/volume concentration. We performed series of simulations calculating the maximum value of MDE for varied value of the filler volume concentration for three types of particles size distribution. Simulations results are compared with experimental data obtained for samples described above (Fig. \ref{fig:vol_conc_dep}). Mass fractions were recalculated to volume fraction.
All three types of size distribution show the same trend as experimental data. The lognormal type has the largest saturation value of MDE. For the small volume concentration the lognormal distribution shows smaller MDE value in comparison with other two types, and starts to grow at higher concentration of magnetic filler. In case of lognormal size distribution the amount of particles is less than for monosized sample due to the presence of larger particles. And since for low concentration the average distance between particles is large, the small number of particles plays significant role and leads to the smaller value of effect.
It was widely reported \cite{049, 050, 051}, that particles tend to form chain-like structures inside the MRE when external magnetic field is applied. We used GTK open library to visualize particles positions throughout the simulation.
Visualization of aligned chains of particles inside the matrix in maximal magnetic field for monosized distribution is shown on figure \ref{fig:mre} and in supplementary video.
\begin{figure}[!htb]
\includegraphics[width=\linewidth]{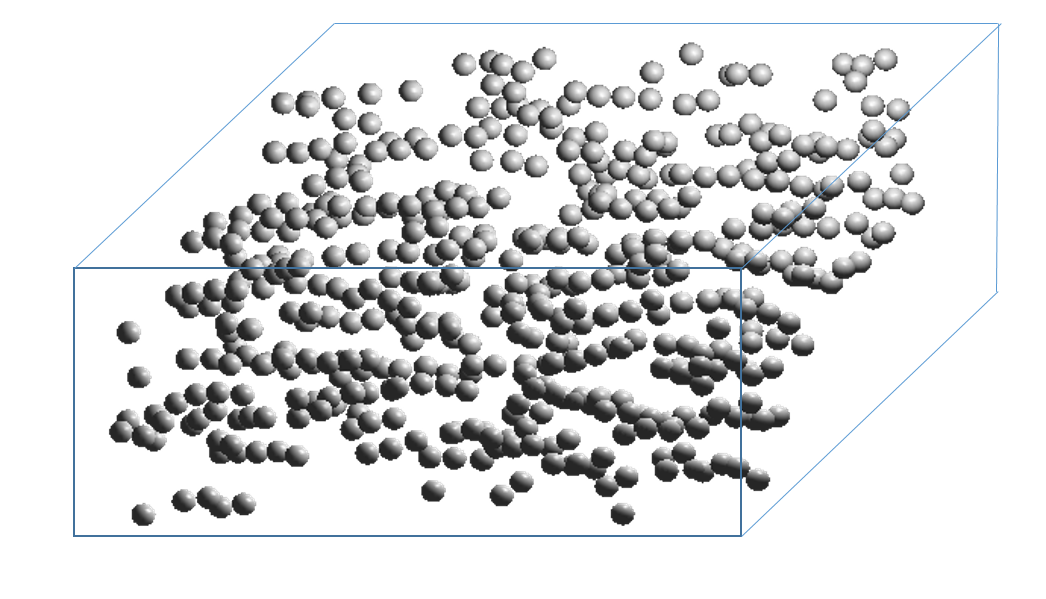}
\caption{Formation of chain-like structures within the elastomer under applied magnetic field.}
\label{fig:mre}
\end{figure}

\subsection{Dependence of MDE on the configuration of the MRE sample}
\begin{figure}[!htb]
\includegraphics[width=\linewidth]{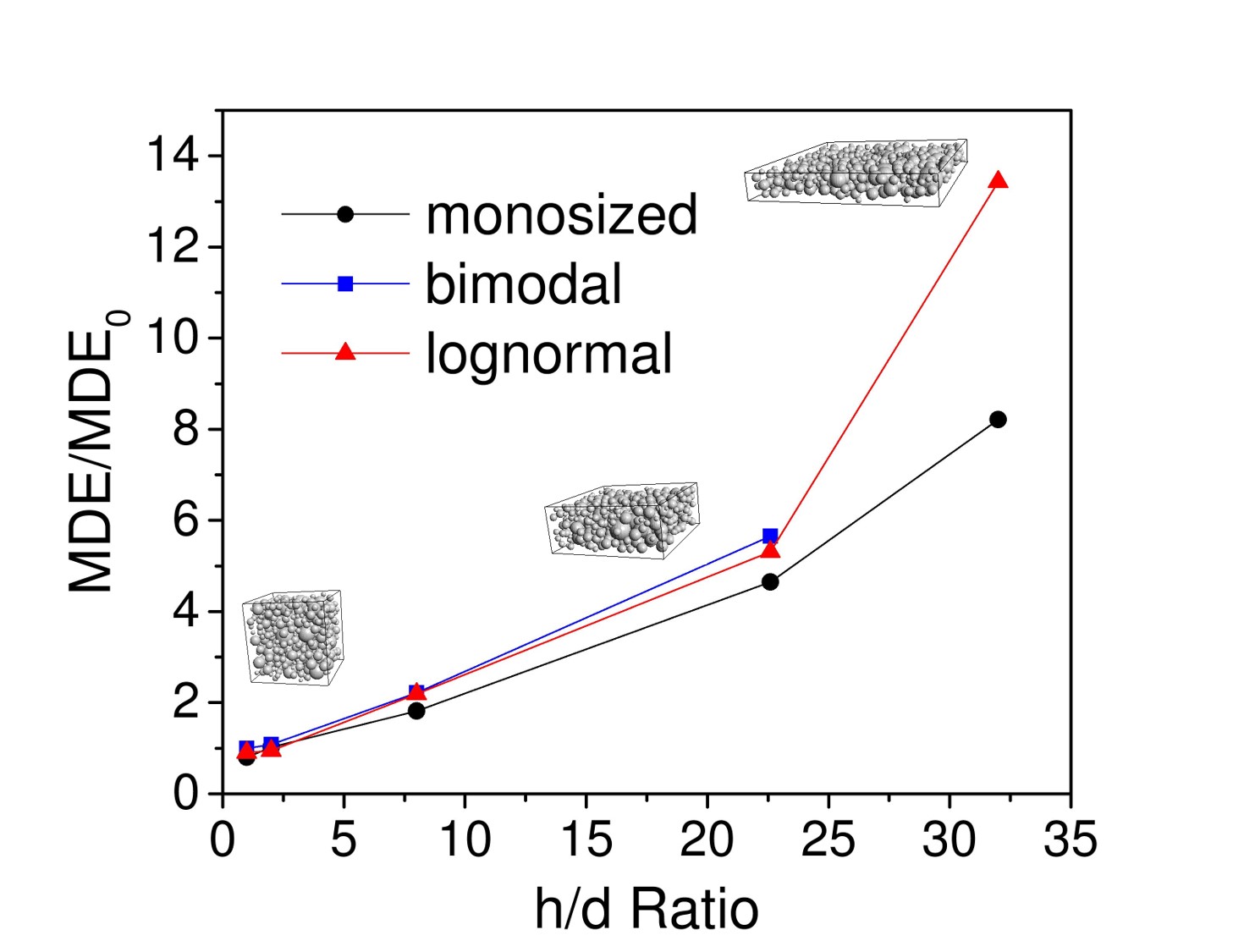}
\caption{Increase of MDE value with increase of h/d ratio of the sample. h - side of capacitor (sample width), d - thickness of the sample. $MDE_0$ is the value of effect for the sample of cubic shape (h/d=1).}
\label{fig:ratio_dep}
\end{figure}
We have performed a series of simulations to obtain the dependence of MDE on the sample shape. We changed width/thickness ratio of the MRE sample and calculated maximum MDE value for each case. Volume concentration of magnetic filler has been fixed and equal to 27\%. Volume of the sample was also fixed. The simulation shows the increase in the effect for the thinner samples. For lognormal type of size distribution the increase was upto 1400\% for ratio = 32 compared to ratio = 1 (Fig. \ref{fig:ratio_dep}). This may be explained by chains formed by particles to be shorter, which leads to increasing influence of the single particle displacement in the chain. Presence of big particles for lognormal type leads to almost two times greater value of MDE than for monosized system.\par

\begin{figure}[!htb]
\includegraphics[width=\linewidth]{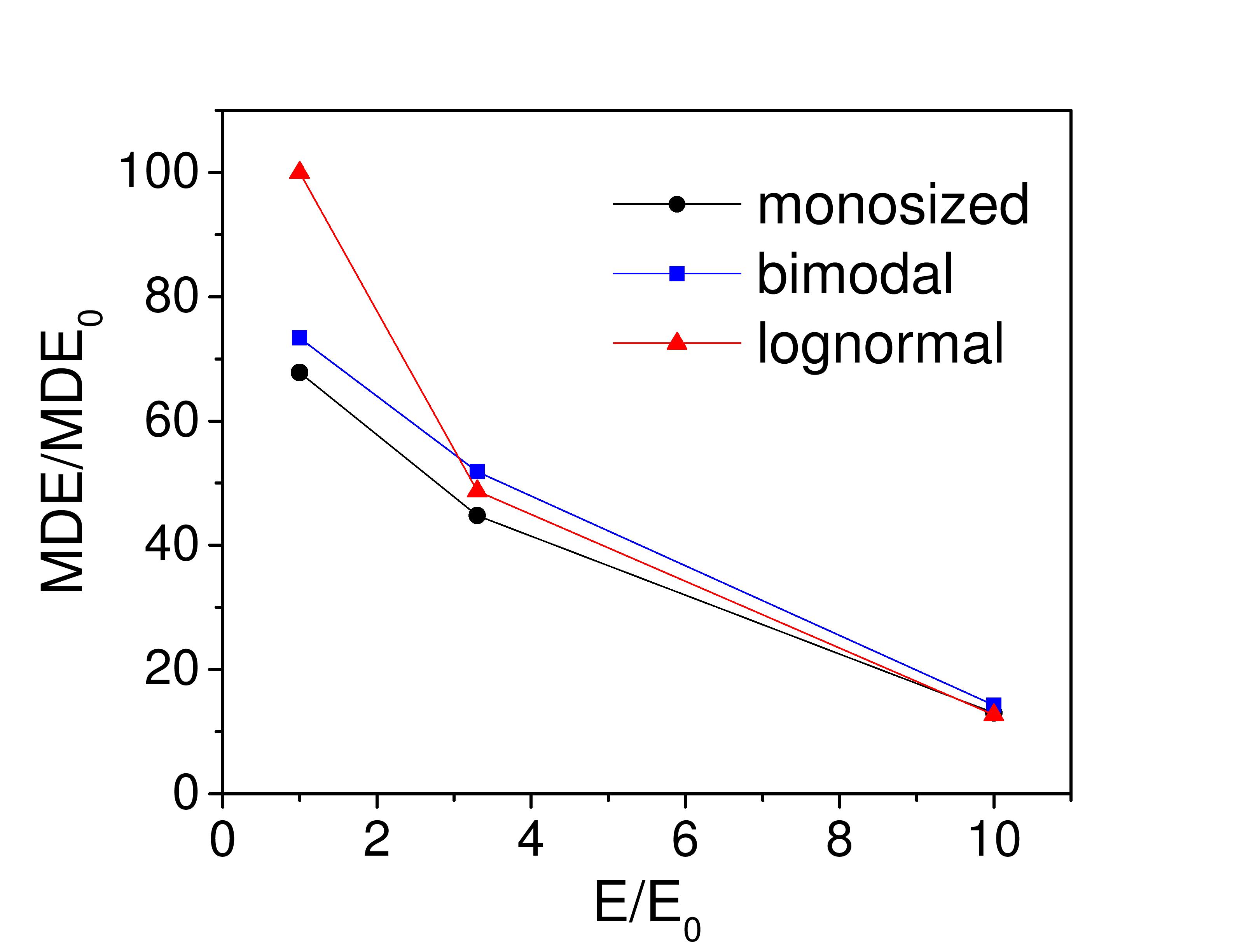}
\caption{Decrease of MDE value with increase of stiffness of the elastomer. $MDE_0$ is the value of MDE for average stiffness $E_0$.}
\label{fig:stiff_dep}
\end{figure}

Also simulations show the decrease of the effect magnitude with the increasing stiffness of samples (Fig. \ref{fig:stiff_dep}). All simulations have been performed for cubic-shaped systems with volume concentration of 27\%. Again, the lognormal distribution type shows the largest value for softer matrices, which is related to the presence of bigger particles in such distribution type. They create larger local magnetic fields and provide the higher mobility which leads to larger MDE value among all three types of size distrubtion.

\section{Conclusions}
In this work we presented the model for simulation of magnetodielectric effect in magnetorheological elastomers and performed the numerical study of the effect for series of samples with different concentration of magnetic filler, size and space distribution of particles as well as elastic properties of polymer matrix. We have found that the effect tends to saturation and has hysteretic behaviour due to the elastic response of matrix. The influence of the orientation of magnetic field was studied as well, the strong anisotropy of effect was observed. Different types of particles size distributions were simulated. Lognormal type was shown to give better qualitative match of the modeling and experimental results than monosized type. Presented model provides good qualitative description of magnetic properties and magnetodielectric effect in magnetorheological elastomers. 

\section{Acknowledgements}
We acknowledge the financial support of RFBR (projects 15-08-99554, 18-32-00354).

\bibliographystyle{elsarticle-num}
\bibliography{2018_isaev_sim_mde_in_mre}

\end{document}